\documentclass{article}
\usepackage{ijcai21}

\usepackage{times}
\usepackage{soul}
\usepackage{url}
\usepackage[hidelinks]{hyperref}
\usepackage[utf8]{inputenc}
\usepackage[small]{caption}
\usepackage{graphicx}
\usepackage{amsmath}
\usepackage{amsthm}
\usepackage{booktabs}
\usepackage{algorithm}
\usepackage{algorithmic}
\usepackage{subfigure}

\makeatletter
\newcommand*\titleheader[1]{\gdef\@titleheader{#1}}
\AtBeginDocument{%
\let\st@red@title\@title
\def\@title{%
\bgroup\normalfont\large\centering\@titleheader\par\egroup
\vskip1.5em\st@red@title}
}

\makeatother
\title{CyGIL: A Cyber Gym for Training Autonomous Agents over Emulated Network Systems}

\titleheader{\emph{IJCAI-21 1\textsuperscript{st} International Workshop on Adaptive Cyber Defense}}

\author{
Li Li$^1$\footnote{Contact Author}\and
Raed Fayad$^2$\and
Adrian Taylor $^{1}$\\
\affiliations
$^1$Defence Research and Development Canada\\
$^2$Dept. of Electrical and Computer Engineering, Queens University, Canada\\
\emails
li.li2@ecn.forces.gc.ca,
raed.fayad@queensu.ca,
adrian.taylor@ecn.forces.gc.ca
}

\begin{document}

\maketitle

\begin{abstract}

Given the success of reinforcement learning (RL) in various domains, it is promising to explore the application of its methods to the development of intelligent and autonomous cyber agents. Enabling this development requires a representative RL training environment. To that end, this work presents CyGIL: an experimental testbed of an emulated RL training environment for network cyber operations. CyGIL uses a stateless environment architecture and incorporates the MITRE ATT\&CK framework to establish a high fidelity training environment, while presenting a sufficiently abstracted interface to enable RL training. Its comprehensive action space and flexible game design allow the agent training to focus on particular advanced persistent threat (APT) profiles, and to incorporate a broad range of potential threats and vulnerabilities. By striking a balance between fidelity and simplicity, it aims to leverage state of the art RL algorithms for application to real-world cyber defence.
\end{abstract}

\section{Introduction}

To strengthen cybersecurity of a networked system, red team exercises are often employed to validate its defensive posture by launching various attacks. Effective red team exercises may incorporate adversary profiles to simulate the most relevant Advanced Persistent Threats (APTs). In certain cases it may also be desirable to test defences against a broad coverage of Tactics, Techniques, and Procedures (TTPs) from various advanced adversaries. 

Red team exercises have significant costs, in both the development of human expertise and in the time required to conduct them. To improve the efficiency of red teaming, tools automating red team exercises such as emulators have grown in popularity, including capabilities like staging frameworks, enabling scripts, and execution payloads \cite{caldera2021,infectionMonkey,atomicRedTeam,empire2018}. These red team automation tools facilitate attack exercises. When using these tools, planning and decision-making, including organizing TTPs through the stages of the campaign,  rely still on the human experts.

The recent rapid advancement of artificial intelligence (AI) \cite{Mnih2013,Mnih2015,Isele2018} brings about the prospect of AI-assisted or even autonomous AI red teaming. Superior dynamic decision-making capability developed through AI learning and training may achieve emerging attack Course Of Actions (COAs) across complex networked cyber systems that are not yet anticipated or developed by the human red team experts. In this approach, the red agents are trained using Reinforcement Learning (RL) and particularly, Deep RL (DRL) algorithms to learn, identify and optimize attack operations in the network. The same analogy applies to the blue agents that may learn and optimize a defence strategy against the red agents. In our preliminary study on red DRL agents using DRL algorithms \cite{MadeenaSultana2021}, both the DQN (Deep-Q-Network) and PPO (Proximal Policy Optimization) agents can be trained to stage and execute optimized network attack sequences despite the uncertain success distributions associated with their attack actions.

The first issue of concern is the training environment. The network cyber environment is much more complex compared with other domains where DRL solutions are being applied. It involves different network segments and many functional layers where the networking devices, protocols, services, applications, users, attackers and defenders are interacting together. The environment is not only entangled with a multitude of dynamic factors and potential actions, but also partially observable with random distributions of action outcomes. 

At present, the few RL/DRL training environments for networked cyber operations~\cite{schwartz2019nasim,Baillie2020,MS2021}, are mostly simulation-based. This is reasonable given the benefits of a simulated environment that models cyber network and its operations using abstractions. This approach is hardware resource-efficient, involves no real networks nor action executions, and as a result expedites training. A potential downside of a simulated environment is its deviation from reality, which may render the trained agent decision model less relevant. Increasing the fidelity of a simulator is difficult because the cyber network operations environment is complex and therefore hard to abstract; the state-space of the simulator increases rapidly when trying to model more details of the real environment.

In comparison, the emulation approach requires many more hardware resources to operate a cyber network and all its assets. The emulated environment is especially expensive in agent training time, because actions are physically executed in the network. Nevertheless the time to learn can still be low compared to a red team exercise which often takes weeks or even months. In addition, the trained AI agents are intended to provide augmented and emerging insights and results beyond those gathered by the human red teams. The key appeal of the emulated training environment is its high fidelity, resulting in an agent more applicable to the real world. The emulated environment also avoids the difficult task of establishing a correct and appropriate model for the complex real system. It can alternatively be used to generate realistic models for building the simulated environment, and to validate the simulated environment models. In general, for practical applications, both simulated and emulated training environments are being used to advise and complement each other.

Implementing an emulation based RL environment is complex and challenging as it involves the real cyber networked system. This work presents CyGIL, an emulated RL/DRL training environment for cyber operations in complex networks. CyGIL takes an efficient system implementation approach by leveraging open source network and cyber operations tools in constructing a modular architecture. CyGIL includes a network emulator but can also run on other emulated or real networks. Though its first prototype supports only red agent training, CyGIL's system architecture is designed for multi-agent blue and red team training, i.e. as a complete cyber-RANGE for AI agents. Embedding the State of the Art (SOTA) red/blue team emulators \cite{caldera2021}, CyGIL inherits all enterprise adversary TTPs per the ATT\&CK framework \cite{attack2021} in the red agent's action space. The action space in different games is then configurable for training agents either focused on specific APTs, or with a broad coverage of emerging APTs and COAs against the network. 

The contributions presented are twofold: a novel architecture approach to achieve an emulated RL/DRL training environment for complex network cyber operations, and the integration of industry SOTA tools to maximize the action abilities of the agents. To our knowledge, CyGIL may be a first emulated RL/DRL environment for networked cyber operations that places the SOTA attack emulation tools to the disposal of AI agents. 

The rest of this paper is organized as follows. Section 2 briefly reviews the current network cyber RL/DRL training environments. The CyGIL system architecture and implementation approach are presented in Section 3. Section 4 illustrates examples of preliminary AI agent training experiments. Section 5 summarizes the concluding remarks and the future research with CyGIL.

\section{Current cyber RL training environments}

The scarcity of RL/DRL training environments for networked cyber operations may be ascribed to the complexity of cyber networks. At present, when applying RL/DRL to network cyber research, the investigation is often confined to a particular cyber task or a network location \cite{nguyen2020deep}. Although a few host based environments for training AI-assisted penetration testers have been reported \cite{pozdniakov2020smart,chaudhary2020automated}, they have very limited game goals and actions. In \cite{Ghanem2018Pentest}, a network pentester training environment is modelled as a Partially Observable Markov Decision Process (POMDP). The details of the environment and the RL training were however not described.

In the OpenAI Gym repository of RL/DRL training environments \cite{openAIGym}, the Network Attack Simulator (NASim) \cite{schwartz2019nasim} supports red agent training for network-wide penetration tests. Being a simulated environment, NASim represents networks and cyber assets, from networking elements like hosts, devices, subnets, and firewalls to services and applications, in abstractions modeled with a finite state machine. The simplified action space includes ``network and host discovery", ``service exploit" for each configured service vulnerability in the network, and ``privilege escalation" for each kind of hijackable process running in the network. The agent can thus simulate a simplified kill chain through discovery, privilege escalation, and service exploits across the network. Some critical network-based adversary abilities like lateral movement are combined into the general `exploit'. At its level of abstraction, NASim does not map concepts of TTPs and APTs for adversary focused attack discovery and optimization. 

Recently Microsoft open-sourced its network cyber RL training environment, the ``CyberBattleSim" (CBS) \cite{MS2021}, also built on the OpenAI Gym. CBS is for red agent training that focuses on the lateral movement phase of a cyber-attack in an environment that simulates a fixed network with configured vulnerabilities. The attacker uses exploits for lateral movement while a pre-defined simple blue agent seeking to detect the attacker and contain the intrusion. Similar to NASim, the CBS environment can define the network layout and the list of vulnerabilities with their associated nodes. In the CBS, the modelled cyber assets capture OS versions with a focus to illustrate how the latest operating systems and up-to-date patches can deliver improved protections. As stated by Microsoft, the CBS environment has a highly abstract nature and cannot be directly applied to real-world systems~\cite{MS2021}.


In \cite{Baillie2020}, a work-in-progress cyber gym called CybORG is presented. CybORG is designed to support both simulated and emulated environment with the same front-end interface. CybORG also supports competing red and blue agents. The simulated environment of CybORG captures certain details of the cyber assets and provides a large action space. For example, the red action space includes exploits from the Metasploit Framework (MSF) space, while blue actions are based on Velociraptor \cite{velociraptor} capabilities. Both red and blue also have access to shell commands for low-level actions. Aiming to be a more realistic network cyber operation environment, the state space of the CybORG simulator is very large. When aiming for this level of detail, an emulator offers an advantage in avoiding the cost of building the requisitely complex state model. To this end, CybORG also supports the emulated environment which is still in testing and development \cite{Baillie2020}. An interesting issue raised by CybORG is the training game design for the environment. The networked cyber environment must support not only many different network configurations but also different games for capturing the potential attack vectors and objectives of the adversary.

Compared with the current solutions, the novelty of CyGIL includes its stateless architecture approach that enables a complex emulated network cyber environment. Its comprehensive action space allows for realistic training games. Additionally CyGIL directly leverages the rapid advancement of industry cyber defence frameworks~\cite{hutchins2011intelligence,attack2021} and automation tools to stay at the forefront of the most relevant APTs and SOTA technologies found in the current cyber network operations.  

\section{CyGil Architecture and System Implementation}
 
As with all RL/DRL training environments, CyGIL trains an agent to optimize its sequential action decisions by maximizing the accumulated rewards collected from its actions. The objective of the agent is modelled by a reward function that returns the reward value after each action execution, when the agent also receives its current observation space (ObS). Different agents have their own action spaces, objectives described by reward functions, and ObSs. As in a real network cyber environment, while the blue agent may see most of the network states depending on the sensors employed, the red agent often starts from seeing nothing in its ObS and builds it up step-by-step from the output of its executed commands. As CyGIL currently is tested only for red agent training, the descriptions hereafter refer only to the red agent perspective unless explicitly stated otherwise.

\subsection{Centralized control and distributed execution}
\begin{figure}[tb]
 \begin{center}
 	\includegraphics[scale=.5]{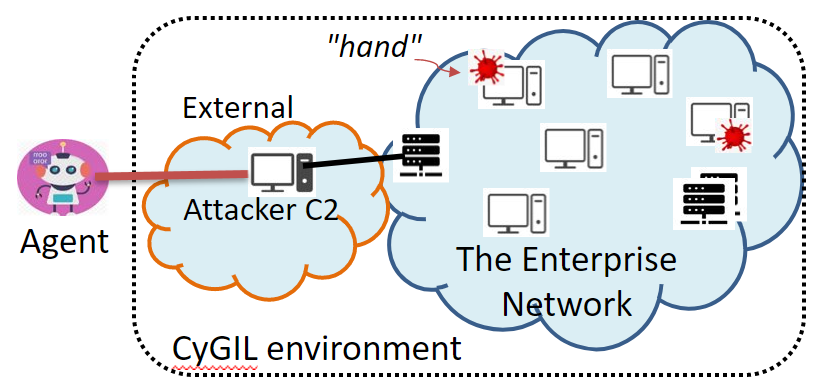}
 \end{center}
\caption
 { \label{fig:C2} 
    The CyGIL Red team C2 Infrastructure - centralized control and distributed execution}
\end{figure}

Cyber operations require suitable infrastructure to deploy and execute. CyGIL employs a command and control (C2) infrastructure, as shown in Figure~\ref{fig:C2} with the red agent example. The centralized C2 decides and launches attack actions which are then executed by the distributed implant(s), denoted as ``hand(s)" in the network. Thus, in CyGIL the red agent that is being trained performs the decision-making C2 for the attack campaign. The agent organizes the action sequence of the attack operation and dispatches to hands the payloads for command execution at different network locations depending on their local configurations. An action at one step of the action sequence may involve multiple executions by different hands at the same time. The ObS of the agent aggregates the information collected by all hands from their executions of the action. The agent's decision model therefore involves the dimension of managing multiple hands to traverse the network for carrying out actions towards the higher reward.  
 
As such the agent executes actions continuously in time in the CyGIL \textit{env}, with each action counted as one action step by the \textit{env}. The attack operation starts after the initial compromise where at least one hand is implanted in the network. The process of the initial compromise, which is often achieved through social engineering (for example email phishing, malicious web link access etc.), is not included in CyGIL.  

\subsection{Functional Substrates}

\begin{figure}[tb]
 \begin{center}
 	\includegraphics[scale=.4]{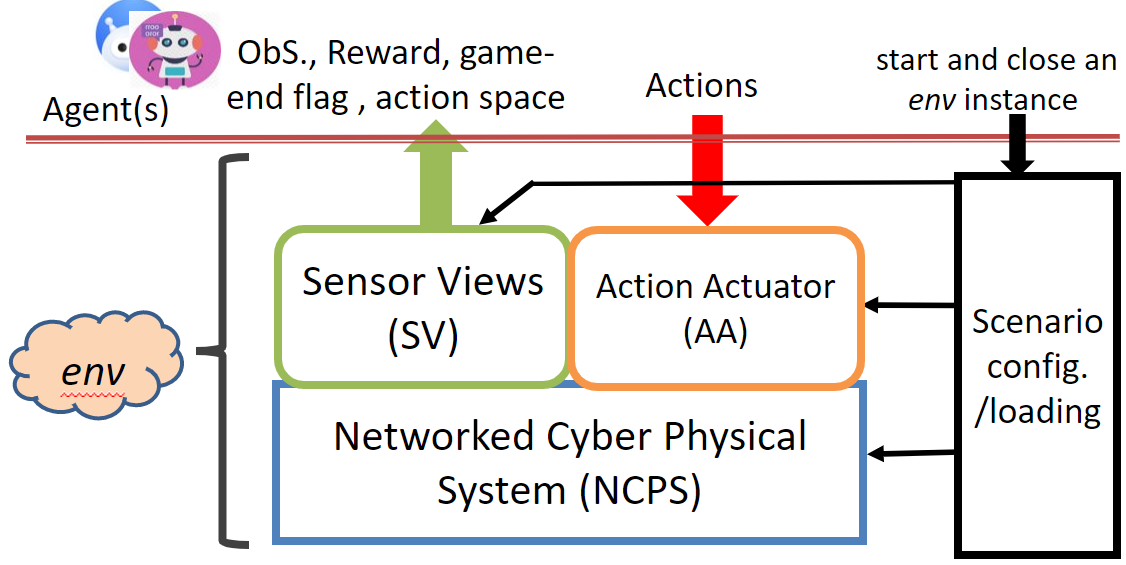}
 \end{center}
\caption
 { \label{fig:cyGil} 
    The CyGIL environment (\textit{env})}
\end{figure}

The CyGIL software architecture is illustrated in Figure~\ref{fig:cyGil}. To enable direct use of all SOTA algorithm libraries published from the RL/DRL research community, CyGIL provides the training environment in the standard OpenAI Gym python class ``\textit{env}". The scenario configuration supports configurations of both the network and the training game. The loading function initializes an \textit{env} instance per the configuration for agent training.  
 
To support the training runtime, the CyGIL \textit{env} embodies three major functional substrates as depicted in Figure~\ref{fig:cyGil}: the Networked Cyber Physical System (NCPS), the Action Actuator (AA) and the Sensor Views (SV). They together wrap the entire cyber network and operational space into the \textit{env} class. Among them, the NCPS comprises all networked cyber assets of hardware and software and enables the SV and AA. The AA carries out the execution of each action on the network through the NCPS. The SV formats and presents to the agents what they can see and know about the \textit{env}.

The NCPS organizes the agent's training on the network in repeated game episodes, each of which is a sequence of the agent's actions against the network. The NCPS verifies if the current training episode is ended in order to reset the \textit{env} for the next training episode. The reset cleans up relics left in the network by the agent and restores the hand(s) to their initial positions. The NCPS keeps for the agent an information database, namely the ``fact" database, to record its executed actions and the results parsed from the actions' output messages, which is used in forming the agent's ObS. 

The AA actuates on the network every action offered to the agent, employing the hand(s), collecting the output messages, and storing the results to the NCPS's information database for exposing the observation space to the agent. The reward resulted from the action execution is also computed according to the reward function.  

Although the action space is supported and actuated by the AA, the SV formats and presents the action space to the agents per the standard interface~\cite{openAIGym} between the \textit{env} and agent. The SV also formats and presents to the agent the reward, ObS and game-end flag after each action execution according to the standard interface. As described above, this information is generated and maintained by the AA and NCPS.  

\subsection{Functional implementation}
The CyGIL implementation builds upon SOTA open source software frameworks and tools, aiming to keep CyGIL relevant with respect to red (and blue) team techniques advanced by the industry and research community. In particular, the CALDERA framework developed by MITRE \cite{caldera2021} was selected as a basis for red actions after evaluating red team emulation platforms for their C2 infrastructure capabilities, plug-in interfaces, and project status. 

Actively developed and maintained by MITRE in open source, CALDERA's purpose is to enable red team emulation using the MITRE ATT\&CK framework that encompasses and analyses reported cyber-attack TTPs. CALDERA thus supports an extensive red team action space. Additionally, the plug-in APIs of CALDERA allow for quick add-on of new custom abilities and other red team tools, e.g. Metasploit, to expand the supported action space. The abilities/actions map to the implementations of the Techniques in the ATT\&CK framework. CALDERA also supports blue agent and red vs. blue "brawl" games. Thus, CALDERA brings to CyGIL a comprehensive and real threat-relevant action space, together with the multi-agent capabilities for red vs. blue competitive games which we plan to leverage in future work.  

\begin{figure}[tb]
 \begin{center}
 	\includegraphics[scale=.485]{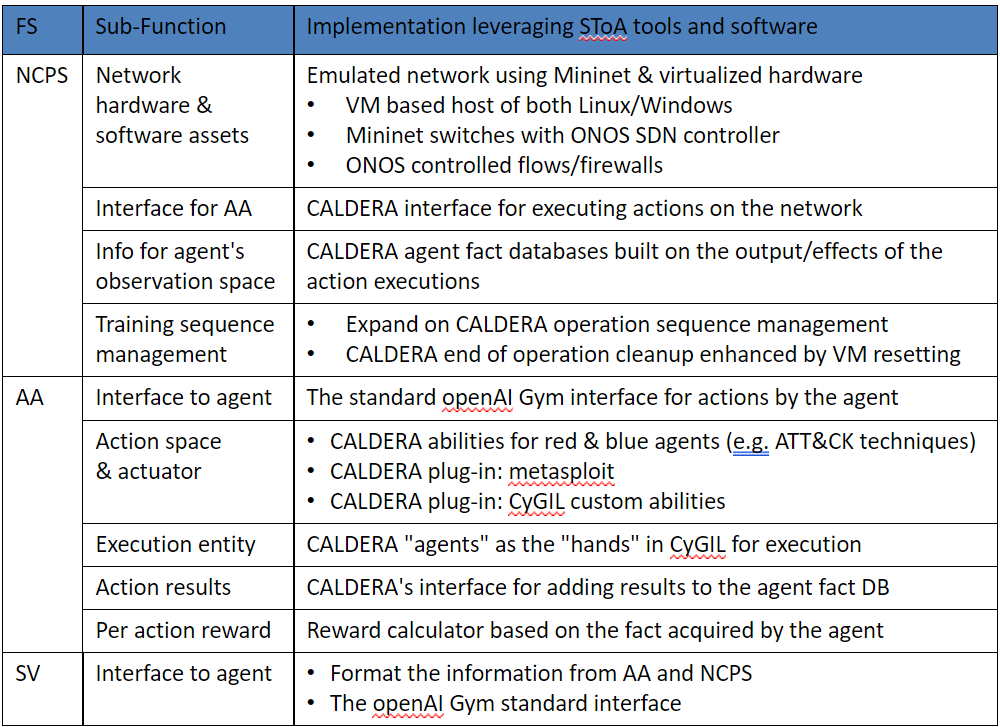}
 \end{center}
\caption
 { \label{fig:implementation} 
    The CyGIL implementation approach}
\end{figure}

As an emulation platform, CALDERA is not a RL/DRL training environment. The CyGIL implementation, as described in Figure~\ref{fig:implementation}, adapts CALDERA to realize the functions required for RL/DRL training \textit{env} with an additional interface layer. Figure~\ref{fig:implementation} also describes other open source tools used to implement the RL/DRL training \textit{env}.

This implementation approach enables CyGIL \textit{env} to provide agents with all the actions offered by CALDERA, and to grow them with new CALDERA releases. The red agent for example has more than 270 actions to apply against Windows systems and more than 180 actions against Linux systems. These actions cover almost the entire ATT\&CK matrix, including execution of various techniques for initial access, discovery, command-and-control, credential access, privilege escalation, persistence, defense evasion, lateral movement, collection, and exfiltration. Additionally, CyGIL enhances the agent action space by integrating Metasploit and adding certain CyGIL custom lateral movement actions through the CALDERA plug-in interfaces.  

\subsection{The red observation space}
After the execution of each action, the ObS must be returned to the agent as part of the result (Figure~\ref{fig:cyGil}). To this end, CyGIL does not maintain the entire network system states for generating the agent's ObS. Instead the NCPS uses CALDERA agent's ``fact" database~\cite{caldera2021} which already gathers the information harvested by the red agent from the output of each action command executed by the hand(s). These are what the red agent realistically may ``see/observe," for example a new host IP address discovered, or the new process IDs obtained that might be used for process hijacking. The information in this fact database in NCPS forms the red agent's ObS, which is inherently partially observable and very limited compared to the state space of the \textit{env}. 

The SV formats the ObS of the \textit{env} to a 2-dimensional array that consists of rows to record found facts of each host and of the entire network. The network row checks if certain network wide information has been gathered, such as the fully qualified domain name, the domain user and admin credentials, the organization names, email servers, etc. Each host row describes a host, where the columns keep the information in the following 10 categories: 
\begin{itemize}
	\item If a hand is on the host and at what user privilege level 	
	\item The local user accounts and creds found	
	\item The OS types and host types if it is a particular server, e.g. the domain controller or a web server
	\item The modifiable service/process  	
	\item The local files, directories, and shares
	\item Host defensive configurations seen, e.g. anti-virus
	\item The host network info found, e.g. domain name, IP address, subnet interfaces
	\item Other host system info found, e.g. system time
	\item Other remote system info found on this host, e.g. the ssh command executed before or the remote IPs seen/used
	\item If the action executed successfully on the host
\end{itemize}

The ObS does not present the exact values or strings of the information but rather applies 0/1 coding as much as possible to indicate if the item is found or not. In a few columns, the number of obtained information pieces is indicated using integers, for example the ``number of local user names found." The content details of the information, for instance the IP address found of a remote host, is not shown in the ObS but is maintained in the CALDERA fact database to be used in forming the commands and payloads for the action selected by the agent before sending it to the hands for execution. It is important to disassociate the agent from the very specific information values that are irrelevant to its decision model. Otherwise, a slight change such as a different username or a digit in the IP address may render the trained agent incapable of generalizing to these trivial differences. 

The number of rows must not reveal the total number of hosts in the network as the agent in reality cannot observe accurately such information. One option is to use a large number of rows to support the maximum number of hosts in a network for which this agent model may be trained. A better alternative is to reduce the returned ObS to only the information gathered by this action execution, excluding all what has been seen by the agent before. With a hand potentially located on each host in the network of the \textit{env}, a composite ObS format such as a list of arrays is needed to achieve such a size reduction.   

Although containing only a small subset of the total state space, the ObS of the red agent can still grow exceedingly fast. Given an ObS array of size $M_r \times M_c $, where $ M_r $ is the number of rows and $ M_c $ number of columns, the size of the observation space $ S_{ob} $ can be expressed as in Equation ~\ref{eq1:complexity}.         
\begin{equation}
	\label{eq1:complexity}
	S_{ob} = (\prod^{N}_{n=2}(n)^{C_n})^{M_r}, \quad s.t.\: \sum^{N}_{n=2}(C_n)=M_c
\end{equation}


where $n$ is the number of possible values in a column, $C_n$ is the number of the columns that take $n$ possible values. Reducing $ N $ towards $N=2$ and reducing $C_n$ and $M_c$ as much as possible will contain $S_{ob}$. We are currently investigating the importance of each column empirically by observing game outputs with the goal of reducing $N$,$C_n$ and $M_c$.  It should be noted that this ObS size does not affect the complexity and running of the CyGIL \textit{env}, while posing challenges to the agent training. Addressing such challenge is the purpose of CyGIL in supporting the cyber AI research.    

\section{Games and Agent Training}

\subsection{CyGIL Mini-testbed Setup}
\begin{figure}[tb]
 \begin{center}
 	\includegraphics[scale=.39]{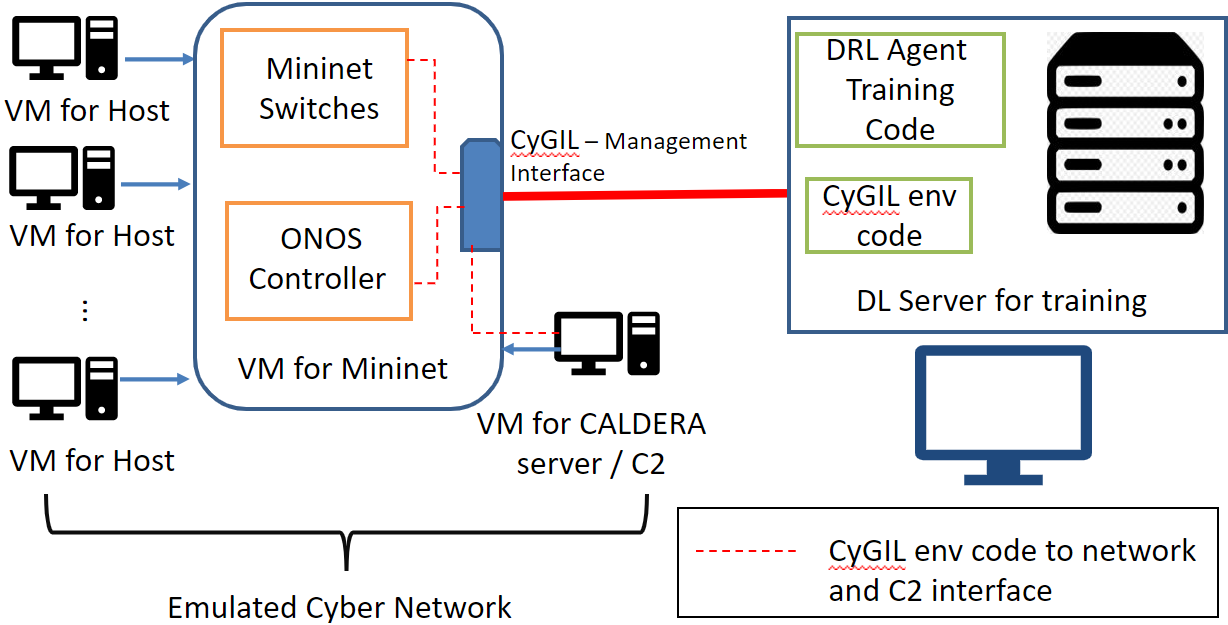}
 \end{center}
\caption
 { \label{fig:system} 
    The CyGIL system setup}
\end{figure}  

A mini-testbed CyGIL environment is shown in Figure~\ref{fig:system}. The emulated cyber network runs on virtualized hardware where the Mininet and ONOS controller manages all the hosts and switches on the target network, with a CALDERA C2 on an external network. The ``CyGIL env code" in the DL server provides the \textit{env} for the agent training. The ``CyGIL env code" initializes the emulated network using Mininet switches and the ONOS controller, and uses the CALDERA server to execute all actions from the agent towards the network. The emulated network can also run on any other network emulator or can even be a real network, as both the CALDERA server and the CyGIL management interface are for connecting to real networks.

In the following experiments, the emulated network is hosted on a Dell Laptop running on Inter(R) Core(TM) i7-4910MQ CPU@2.90Ghz with 32GB RAM. The DL training server in Figure~\ref{fig:system} is a HP laptop running on Inter(R) Core(TM) i9-98880H CPU@2.30Ghz with 64GB RAM and a NVIDIA Quadro RTX 5000 Graphical Processing Unit. The OpenAI Gym standard interface of CyGIL \textit{env} enables support for almost all open source SOTA DRL algorithm libraries. Trainings are tested on both the Tensorflow (version 2.3.0) and Pytorch (version 1.7.0) DL frameworks.  

\subsection{Experiments}
A CyGIL \textit{env} instance includes both the network and game configurations. Different games can be set up over the same network to form different \textit{env} instances for training agents of different profiles in capabilities and objectives. CyGIL allows flexible game configurations including the selected action space, the variables in the reward function, the game goal, the maximum number of actions allowed per game episode, and the initial position(s) of the hand(s). The details of the CyGIL game definition framework is a very important topic but beyond the scope of this paper. In the game examples presented in this work, the reward is simply set as $-1$ per executing hand for all actions, except the successful last action that reaches the game goal, whose reward is $ 99 $. The positive reward placed only on the last ``goal action" pushes the agent to reach the end goal in as few action steps and with as few hands as possible. It also prevents the human game designers from subjectively influencing the agent's COAs. Note that CyGIL allows more types of game setup in reward functions, action cost, and game-end objectives than shown in the examples presented here. In the following, a training episode ends when either the maximum number of actions is reached, or when the goal action succeeds, that is, the agent's reward reaches the maximum value. Then the \textit{env} is reset for a new episode. 

When selecting a small subset from the total action space in a game, the training may focus on the most relevant TTPs associated with the selected adversary profiles. A larger action space may assume an adversary with more or even all the supported TTP abilities, generating complex games that require longer training durations. The ample action space in CyGIL enables the red agent to execute all the typical tactics following the initial network breach to delivery in an end-to-end kill-chain operation, e.g. achieving persistence and lateral movement in the network to collect and exfiltrate target information.    

\begin{figure}[tb]
\centering     
\subfigure[Network scenario]{\label{fig:a}\includegraphics[width=50mm]{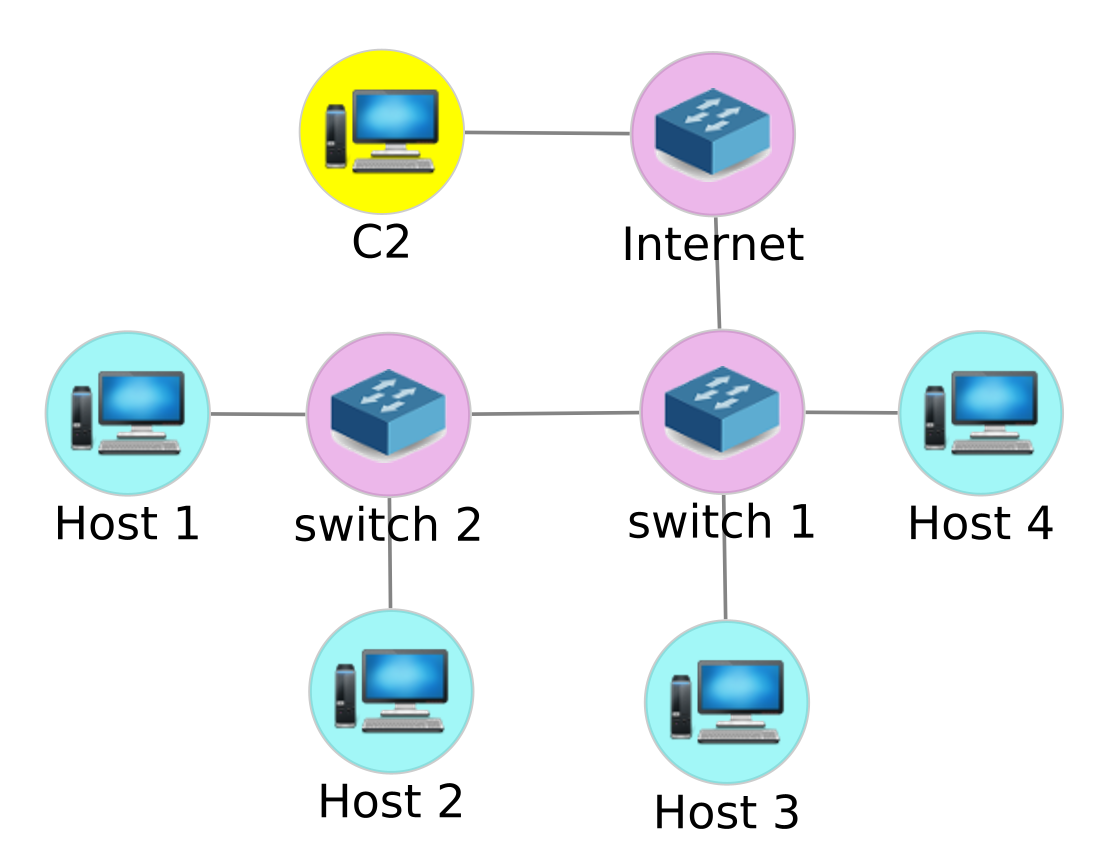}}
\subfigure[Action space]{\label{fig:b}\includegraphics[width=70mm]{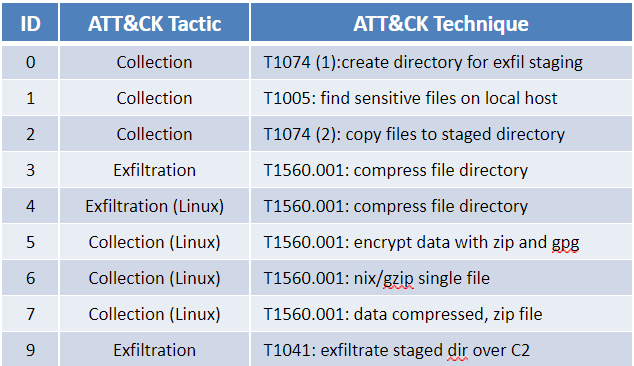}}
\caption{Training game example 1}
\label{fig:game1}
\end{figure}

The first simple testing game is illustrated in Figure~\ref{fig:game1}. The action that succeeds in exfiltrating the target file from the host to the C2 gets the reward of $ 99 $ and ends the game episode. The maximum number of actions in one training episode is 100. The action space (Figure~\ref{fig:game1}(b)) includes variants of techniques from two groups of adversary tactics, namely collection and exfiltration. Depending on configurations of the hosts, the exploits may or may not be executable. Even if executable, the exploit may or may not succeed due to the host states. In the network (Figure~\ref{fig:game1}(a)) host 1 to host 3 are Windows 10 machines and host 4 is on Linux Ubuntu. Data traffic is sent between the following host pairs: host 1 and host 3, host 1 and host 2, and host 3 and 4. 

\begin{figure}[tb]
\centering     
\subfigure[Loss]{\label{fig:game1Resultsa}\includegraphics[width=40mm]{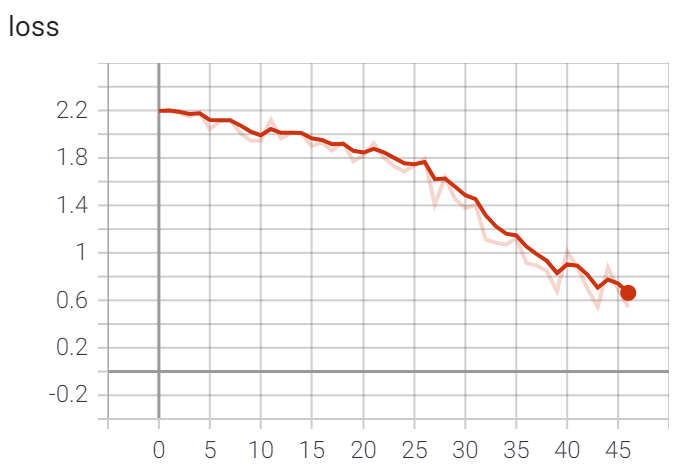}}
\subfigure[Reward]{\label{fig:game1Resultsb}\includegraphics[width=40mm]{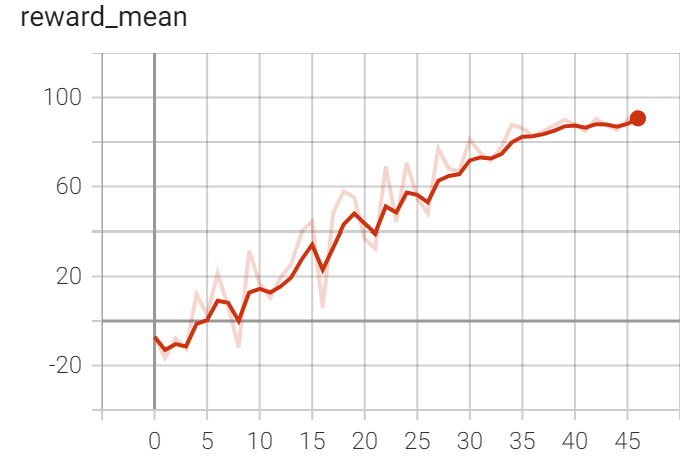}}
\caption{Training the DQN agent in game 1 - x Axis in 100 steps per unit }
\label{fig:game1ResultsDQN}
\end{figure}

Two typical DRL algorithms have been used in this test. The first is the Deep Q-Network (DQN) \cite{Mnih2013} using its implementation from the reinforcement learning library Tf-Agent 2.3 \cite{TFAgents} developed by Google on Tensorflow. The second is a classic cross-entropy (CE) algorithm implemented on the PyTorch framework. While the DQN trains on every step, the policy-based CE trains on each complete game episode. The CE is a less powerful algorithm and takes more time to train, which results in a stress-test of the CyGIL \textit{env}. With both algorithms, the agent can indeed learn the optimized action sequence to exfiltrate target files solely through playing in the CyGIL \textit{env}, starting with no knowledge including hosts or actions. The DQN training results are depicted in Figure~\ref{fig:game1ResultsDQN}. The maximum reward of 96 with a sequence of 4 steps using a single hand is reached by the agent after playing 4600 steps. 

\begin{figure}[tb]
\centering     
\subfigure[Network scenario]{\label{fig:game2Net}\includegraphics[width=60mm]{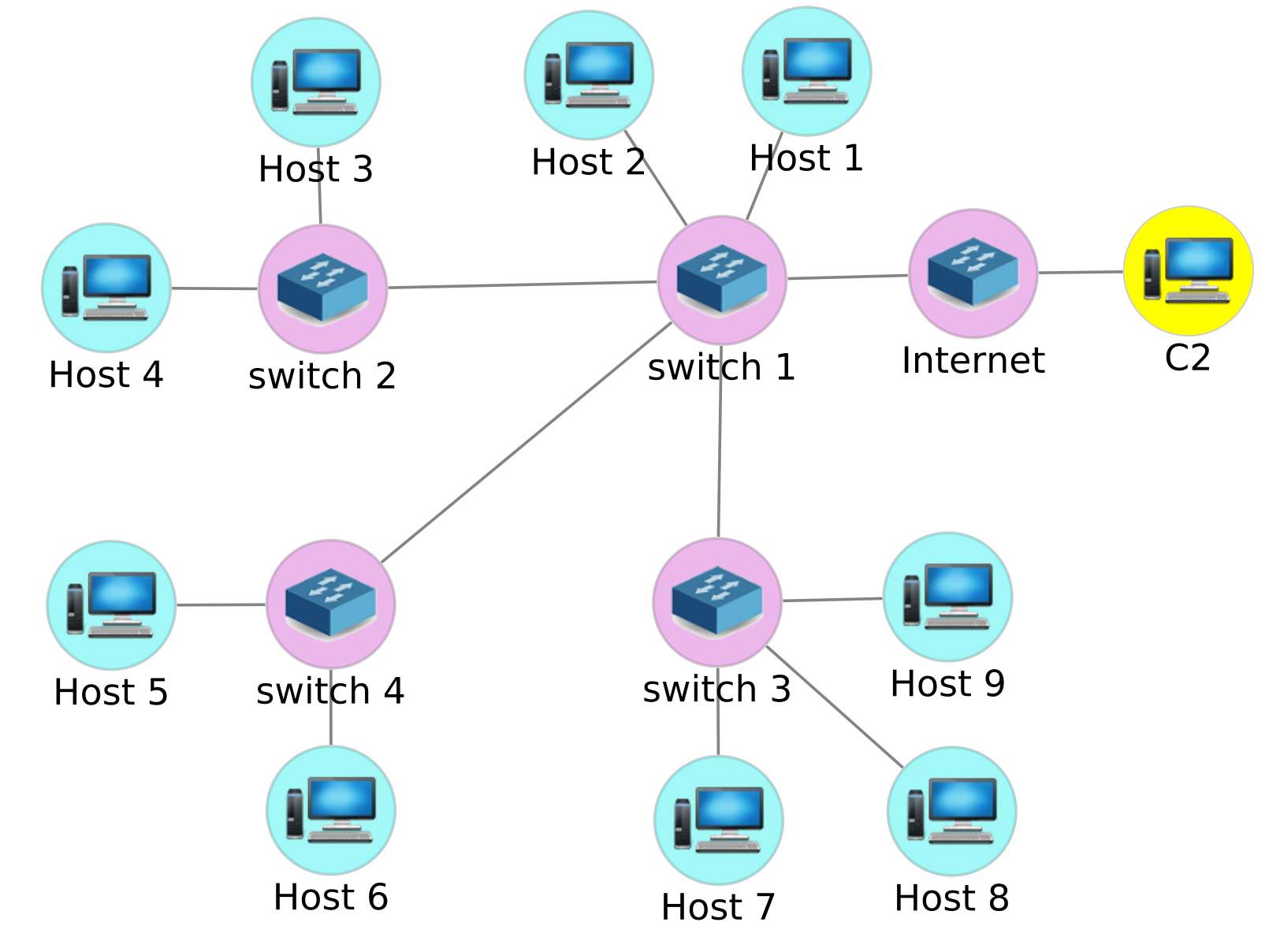}}
\subfigure[Action space]{\label{fig:game2Action}\includegraphics[width=85mm]{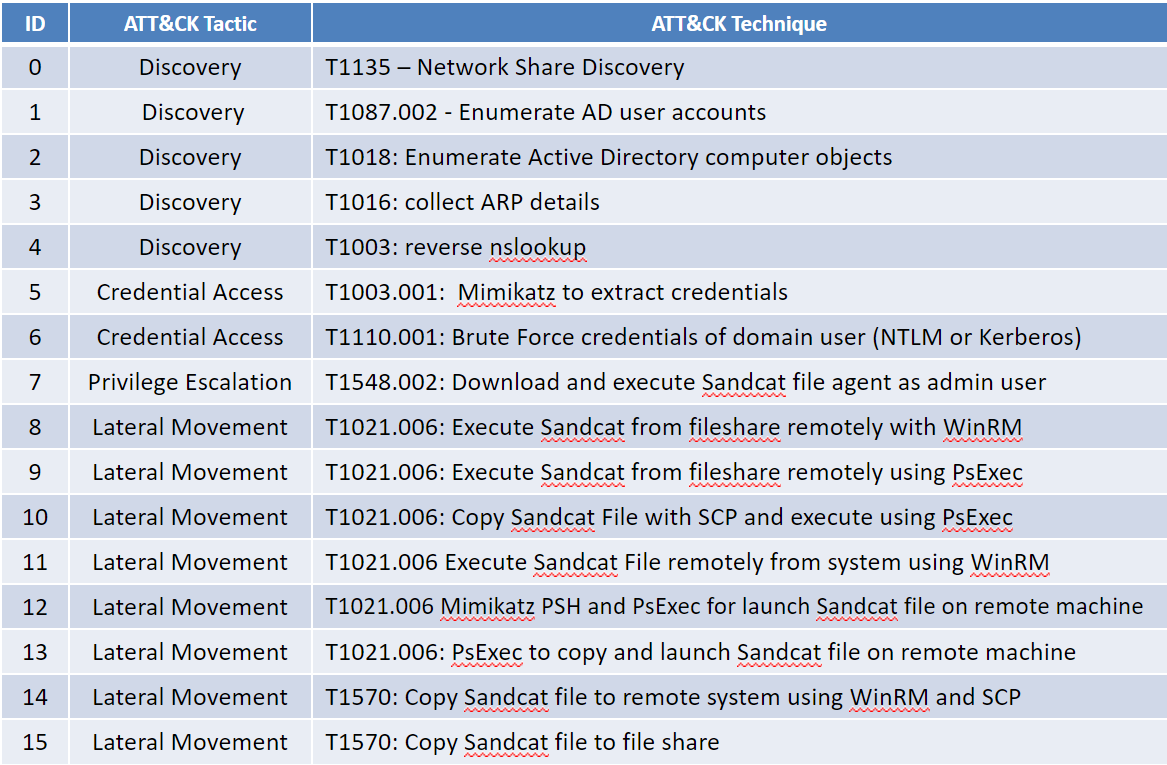}}
\caption{Training game example 2}
\label{fig:game2}
\end{figure}

The second game is on the network illustrated in Figure~\ref{fig:game2} (a), which consists of 9 hosts. The action space is shown in Figure~\ref{fig:game2}(b). In the network, hosts 2 and 5 are on Ubuntu Linux and host 9 is a Windows 2016 server. The rest of the hosts run on Windows 10 Enterprise OS. Hosts 1 and 2 are reachable from the external ``Internet'' by the C2. All hosts inside the network can reach the Active Directory Server/Domain Controller (DC) at host 9. Hosts on the same switch belong to the same subnet and can communicate with each other. Between different subnets, firewall rules are put in place through ONOS to allow host 6 to communicate with host 2 and host 3. In the training, each host sends light traffic to at least one other host as the generic network user traffic. 
  
In Game 2, the initial breach of a single hand/implant escalates to a major network compromise. The action space provides various techniques in four tactic groups of discovery, credential access, privilege escalation and lateral movement, which are used often by high profile APTs. The goal action that succeeds to land a hand on the DC (host 9) receives the reward of 99 and ends the episode. A hand must have an escalated privilege of domain administrator in order to land on the DC, representing potentially a major system breach. The maximum number of action steps per game is set at 300.  

The hand initially lands on host 2. The actions have different conditional success distributions on different hosts and under different circumstances, even for carrying out the same technique. The success rates of actions in this game are relatively low, especially for those of lateral movement. The agent needs to learn the network reachability paths, required credentials, and privilege levels and where, when, and how to obtain them. It also has to manage the best number and placement of hands and learn to select actions among implementation variations of the same technique for higher success rates at different network locations.

\begin{figure}[tb]
 \begin{center}   
	\includegraphics[width=60mm]{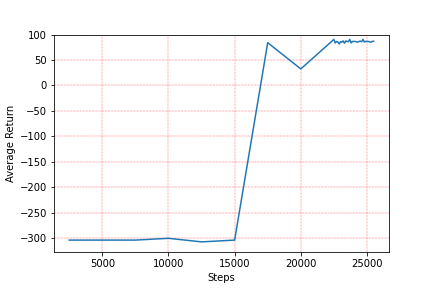}
 \end{center}
\caption
	{\label{fig:game2ResultsDQN}
	 Training DQN agent in game 2}
\end{figure}

Game 2 is drastically more complicated and takes more training time than game 1. The optimized COA's see the agent uses the hand on host 2 via taking on actions 3,4,7 and 13 (or 3,7,4 and 13, or 7,3,4 and 13) to discover, to elevate the privilege and to launch a new hand on host 6. Capturing the domain admin credential on host 6 via action 5, a hand finally succeeds in logging into (via action 12) the DC's domain admin account on the Active Directory server residing at host 9. Using the DQN from Google's Tf-Agent library, the agent through training learns the optimized COA's and reaches rewards ranging from 90 to 92 as depicted in Figure~\ref{fig:game2ResultsDQN}, as some actions have low success ratio. More efficient algorithms are however desirable to improve the training efficiency. This more realistic game presents a challenging problem in DRL training, which is exactly the purpose of CyGIL.    

Action execution in CyGIL runs on wall-time and results in longer latency than in simulated environments. The latency depends on the hardware resources offered to the emulated network. At present in the mini-testbed, average action execution time is under 1s in game 1, and ranges from under 1s to a couple of minutes for actions in game 2. Some actions, such as lateral movement, may take up to 120s for the new hand to install on the remote host and its reporting beacons to be heard by the C2, under the network traffic conditions. The long latency is partly due to the hardware limitation when running more VMs in game 2 on one laptop. Each training episode lasts for about 30s on average in game 1 and about 20-30 minutes in game 2. The episode decreases in its duration when more optimized action sequences are formed that involve fewer steps. The agent training time takes from several hours to a few days, depending on the game and the algorithm. In addition to the action execution latency, the time required to reset the \textit{env}  for a new episode lasts from 3s in game 1 to 40s in game 2 with the mini-version of the hardware platform, contributing also to the average episode duration. To reduce latency both in action execution and game episode/\textit{env} resetting, improved hardware resource including servers and the hyperconverged management infrastructure of VxRail are being tested for network emulation. While impeding the training time, the inherent latency in an emulated environment nevertheless contributes to its realism because as in the real cyber networks, the agent must adapt to variable communication delays between the hands and the C2/agent.

\section{Conclusions and Future Work}

As presented, CyGIL is an emulated RL/DRL training environment for network cyber agents. The CyGIL approach enables industry SOTA red team tooling abilities to support a comprehensive action space. The CyGIL functional architecture realizes a stateless implementation of the \textit{env} that is scalable and expandable. Being an emulated environment, the agent trains on real network configurations to construct relevant and applicable cyber operation decision models. CyGIL provides flexible network scenarios and game configurations to support different adversary objectives and ability profiles. Compliant to the OpenAI Gym standard interface, almost if not all the SOTA RL/DRL algorithm libraries developed by the research community can be used directly with CyGIL. Some preliminary experiments have produced expected training results and highlighted RL/DRL algorithm challenges when applied to the autonomous cyber operation domain.

In our current research, the CyGIL testbed is being improved to reduce latency in all areas, especially with game/\textit{env} resetting. In addition to scaling up the hardware resource, the option of switching between parallel emulated networks for the \textit{env} is being evaluated. At the same time we are investigating how to map the cyber operation goals to relevant training games, as well as the effective DRL algorithms that reduce the training time and improve the model generalization. Finally we are adding blue agent capability to enable competitive multi-agent training for realizing a fully representative red vs. blue cyber game environment.

\bibliographystyle{named}
\bibliography{ijcai21}

\end{document}